\documentclass[aps,prb,reprint,showpacs,superscriptaddress]{revtex4-2}
\usepackage{amsmath}
\usepackage{amssymb}
\usepackage{amsthm}
\usepackage{bbm}
\usepackage{graphicx}
\usepackage{hyperref}
\usepackage{physics}
\usepackage{systeme}
\usepackage{times}
\usepackage{epstopdf}
\usepackage{float}

\hypersetup{
  colorlinks=true,linkcolor=blue,citecolor=blue,
  filecolor=blue,urlcolor=blue,breaklinks=true
}

\begin{document}

\title{Quantum particle motion on the surface of a helicoid in the presence of harmonic oscillator}

\author{Marcos C. R. Ribeiro Jr.}
\email{marcoscezarrrj@gmail.com}
\affiliation{
        Departamento de F\'{i}sica,
        Universidade Federal do Maranh\~{a}o,
        65085-580, S\~{a}o Lu\'{i}s, Maranh\~{a}o, Brazil
      }

\author{M\'{a}rcio M. Cunha}
\email{marciomc05@gmail.com}
\affiliation{
        Departamento de F\'{i}sica,
        Universidade Federal do Maranh\~{a}o,
        65085-580, S\~{a}o Lu\'{i}s, Maranh\~{a}o, Brazil
      }

\author{Cleverson Filgueiras}
\email{cleverson.filgueiras@dfi.ufla.br}
\affiliation{
        Departamento de F\'{i}sica,
        Universidade Federal de Lavras, Caixa Postal 3037,
        37200-000, Lavras, Minas Gerais, Brazil
      }

\author{Edilberto O. Silva}
\email{edilbertoo@gmail.com}
\affiliation{
        Departamento de F\'{i}sica,
        Universidade Federal do Maranh\~{a}o,
        65085-580, S\~{a}o Lu\'{i}s, Maranh\~{a}o, Brazil
      }

\date{\today }

\begin{abstract}
The geometric potential in quantum mechanics has been attracted attention recently, providing a formalism to investigate the influence of curvature in the context of low-dimensional systems. In this paper, we study the consequences of a helicoidal geometry in the Schr\"{o}dinger equation dealing with an anisotropic mass tensor. In particular, we solve the problem of an harmonic oscillator in this scenario. We determine the eigenfunctions in terms of Confluent Heun Functions and compute the respective energy levels. The system exhibit several different behaviors, depending on the adjustment on the mass components.
\end{abstract}

\maketitle

\section{Introduction}
\label{intro}

Currently, the development of a large number of new materials and technologies are possible due the tools provided by Quantum Theory \cite{hills2019modern,liu2009carbon,chaudhary2013graphene,moore2010birth,kushwaha2013carbon}. For example, the study of the Physics of materials such as Graphene, Carbon Nanotubes and Topological Insulators \cite{RevModPhys.83.1057} are very interesting. These materials have several physical properties that allows important applications.  In fact, their optical, magnetic and transport properties has been an active and wide field of research in Condensed Matter Physics, experimentally and also at the theoretical level \cite{laird2015quantum,cheng2019guiding,zhang2008tuning,sevinccli2008electronic,PhysRevLett.79.5086}.

Graphene is an example of two-dimensional material.  Low-dimensional systems have been attracted the interest of the scientific community because of their possibilities in applications such as more efficient electronic devices \cite{yang2019perspective}, medical applications \cite{rezapour2017graphene} and water treatment \cite{liu2013application}, for example. Nowadays, the experimental fabrication of several types of low-dimensional materials is a reality in laboratories around the world \cite{balkanski2012fabrication}. Also, the current technology turns possible the creation of samples in very small scales possessing a single-atom width
 \cite{olsen2018discovering,ferrari2015science,kou2017two,koppens2014photodetectors}.

On the other hand, the intersection between Physics and Geometry produces fruitful results. The description of a classical particle motion, for example, depends on which geometry the particle is immersed \cite{d1988canonical}. The Geometric Optics is another example of relevance of Geometry in describing physical phenomena. In the context of modern physics, the General Relativity Theory \cite{d1992introducing} requires geometrical quantities in its framework. Thus, Geometry plays a fundamental role in research areas such as Cosmology and Gravitation.
Geometrical effects also are relevant in Quantum Mechanics, and in its applications in the investigation of systems in the domain of the Solid State Physics. It is possible, for instance, to investigate the geometrical influence in the dynamics of an electron in a given space \cite{dandoloff2010geometry,dandoloff2009quantum,atanasov2008curvature}.
Also, it is possible to study the influence of topological defects in quantum mechanics. In a remarkable work, Katanaev and Volovich \cite{katanaev1992theory} have showed that the same geometric tools employed in General Relativity could be successful applied to study defects in solids.
It allows to establish bridges between quantum mechanics, condensed matter systems and other areas. For example, condensed matter systems can be used as laboratory for gravitation and cosmology \cite{MORAES2000}.

In nonrelativistic quantum mechanics, the Schr\"{o}dinger equation describes the dynamics of a particle in the presence of a given potential. Since this equation can be used in applications, an interesting theoretical development it is related to the inclusion of geometric effects into the equation, in order to describe a low-dimensional system in the presence of curvature, for example. In this context, important contributions were given by Jensen \cite{JENSEN1971586} and Da Costa \cite{da1981quantum,PhysRevA.25.2893}. More specifically, they introduced a new approach to describe the dynamics of two-dimensional system in a curved surface, immersed in the Euclidean space $R^{3}$.
Da costa's approach initially consider a particle in a three-dimensional space. Despite of that, if the particle is constrained to lie only on a two-dimensional region, it is demonstrated that a new kind of potential emerges: a geometric potential one presenting dependence on the mean and gaussian curvatures of the surface considered \cite{da1981quantum,atanasov2008curvature}. This model is suitable in the context of thin-layers.
The study of the geometrical potential and its implications in quantum phenomena are an active branch of research. The Schr\"{o}dinger equation considering the geometric potential in the presence of electromagnetic potentials was derived in \cite{ferrari2008schrodinger}. An experimental realization of a optical analogue of the geometric potential was reported in \cite{PhysRevLett.104.150403}.
We give more details about Da Costa's model in section 3.
Quantum systems in curved spaces \cite{dewitt1957dynamical,audretsch2012quantum} also are explored in the literature under other points of view. For instance, the problem of a hydrogen atom in a spacetime with a topological defect was considered in \cite{PhysRevD.66.105011}. In \cite{PhysRevD.90.045041}, it is investigated how some quantum communications protocols are affected when they are performed in a curved spacetime.

An interesting aspect in Condensed Matter systems refers to the idea of effective mass: It is possible to describe the behavior of an electron in a periodic potentials by employing an effective Schrödinger equation using a effective mass $M^{*}$ instead of the usual electron mass \cite{burt1992justification}.
The effective mass also can be different in different regions, presenting a anisotropic behaviour. In this context, a possibility of investigation refers to the study of the Schr\"{o}dinger for a particle with a given effective mass considering a curved space, describing, for instance, electronic states of curved samples.

Some particular geometries are interesting since they occur in nature. The helicoidal geometry naturally occurs in Chiral Liquid Crystals \cite{PhysRevE.100.052703}, in macromolecules of DNA \cite{yakushevich2006nonlinear,barbi1999helicoidal} and in fibril structures in animals and plants \cite{ribbans2016bioinspired}.
The helicoidal geometry has been studied in several scenarios in physics, like in the context of branes and black-hole physics \cite{armas2015new}, for example. In Optics, it was demonstrated that linearly polarized light traveling in a helicoidal optic fiber can acquire a Berry's geometric phase \cite{Wassmann:98}.
A study dealing with the electronic states near the Dirac points in helicoidal graphene was reported in \cite{PhysRevB.92.205425}. In \cite{zhan2017graphene}, it was proposed a nanospring consisting of a graphene nanoribbon-based helicoid structure.
Also, an analog of the Hall effect can be induced by an effective electric field in the helicoidal geometry, making possible a charge splitting even in the absence of electromagnetic fields \cite{atanasov2009geometry}.
Recently, Souza et al. \cite{souza2018curved} have studied the behavior of a noninteracting two-dimensional electron gas with anisotropic mass considering several geometries, including the helicoidal one.

In this paper, we consider the problem of a quantum harmonic oscillator in the scenario of an helicoidal geometry and anisotropic mass. More specifically, we solve the Schr\"{o}dinger equation for a particle in a curved space consisting of a helicoidal ribbon, taking into account the corresponding geometric potential and also an harmonic oscillator potential. Our interest in studying this subject is due the fact that the harmonic oscillator is one of most fundamental systems in physics, and a essential model in several applications. Thus, a relevant issue consists in studying this system on different geometries.

The paper is organized as the following: Section 2 is dedicated to the idea of anisotropic mass. In addition, we consider the main aspects of the Schr\"{o}dinger equation with a generic geometric potential with anisotropic mass. Also on section 2, we deal with the problem of a quantum particle on a helicoid and anisotropic mass. Section 3 is dedicated to the problem of a quantum particle subjected to an harmonic oscillator potential, in a helicoidal geometry and anisotropic mass.  We solve the corresponding wave equation and evaluate the eigenstates and determine the ground-state and the first excited state energies. The solution is given in terms of Confluent Heun Functions.
In section 4, we make our concluding remarks.

\section{The anisotropic effective mass and Schr\"{o}dinger equation in a curved space }

\label{sec2}
In this section, we briefly discuss how we can incorporate anisotropic mass formalism in the Schr\"{o}dinger
equation. First of all, it is worth to note there are two possibilities in studying the Schr\"{o}dinger
equation in the context of anisotropic effective mass. We can consider that the effective mass is a tensor and their components are constants. The other possible approach consists of taking it as a position-dependent function \cite{sebawe2016exact}. In this paper, we consider the first one. We are not interested in considering a position-dependent mass at this point. Instead, we want to describe a two-dimensional structure considering two different effective masses: the first one is related to the surface itself, while the second one is related to the normal degrees of freedom. We will give more details in the next section.
The non-relativistic Hamiltonian describing the dynamics of electrons in semiconductors structures, considering the approximation of effective mass, is given by
\begin{equation}
\hat{H}=-\frac{\hbar^2}{2}\left[\frac{1}{M^*}\right]^{ij}\partial_i\partial_j
+V, \label{ef-mass-schro}
\end{equation}
satisfying the eigenvalue equation for the energy levels $\hat{H}\Psi=E\Psi.$ The quantity  $[1/M^*]^{ij}$ is the effective mass tensor. We will take the following diagonal form \cite{anisotropicmass}:
\begin{equation}
\left[\frac{1}{M^*}\right]^{ij}=\left(\frac{1}{\hbar^2}\frac{\partial^2 E}{
\partial k_i\partial k_j}\right)_{\mathbf{k}=0}=
\begin{pmatrix}
M_{11}^{-1} & 0 & 0 \\
0 & M_{22}^{-1} & 0 \\
0 & 0 & M_{33}^{-1}
\end{pmatrix}.  \label{emass}
\end{equation}
The elements $M_{11}$, $M_{22}$ and $M_{33}$ can be different from each other.

Hereafter, we specialize in the case $M_{11}=M_{22}\equiv M_1$ and $
M_{33}\equiv M_2$. Now, we can generalize it, by including the information of a curved space into the Hamiltonian. In three dimensions, the Schr\"{o}dinger equation (without electromagnetic potentials) in a curvilinear coordinate system is given by
\begin{equation}  \label{eq:schr3}
i\hbar\partial_t \psi= -\frac{\hbar^2}{2}\left[\frac{1}{M^*}\right]
^{i^{\prime}j^{\prime}} \left[ \frac{1}{\sqrt{G}}\partial_i\left(\sqrt{G}
G^{ij}\partial_j\right)\psi\right].
\end{equation}
Here, $G=\det(G_{ij}) $, being $G_{ij}$ the metric tensor and $G^{ij}$ is its inverse. In this equation, we are considering the Einstein summation convention whose index are $i^{\prime},j^{\prime}=1,2,3$. Then, we are ready to revise the Da Costa's approach to describe the dynamics of a quantum particle confined in a two-dimensional surface. Imagine a two-dimensional surface S. We can use parametric equations given by $\mathbf{r}=\mathbf{r}(q_{1},q_{2})$ to describe S. Here, $\mathbf{r}$ is the vector that indicates the position of any point of S. Imagine the surface which is immersed in a three-dimensional space. If an arbitrary point it is in the neighborhood of the surface S, we can localize it by employing a vector given by
$\mathbf{r}(q_{1},q_{2})+q_{3}\,\mathbf{n}(q_{1},q_{2})$, which consists of a combination of the vector $\mathbf{r}(q_{1},q_{2})$ on the surface and a vector $q_{3}\,\mathbf{n}(q_{1},q_{2})$, normal to the surface. More specifically, $\mathbf{n}$ and $q_{3}$ are the unit vector and the corresponding coordinate in the normal direction, respectively.
Thus, in the following, we consider the indexes for the surface assuming the values 1, 2, while the value 3 will be related to the normal direction. We can write a relation between the metric tensor $G_{ij}$ in the three-dimensional space near to the surface S and the two-dimensional metric tensor of the surface $g_{ab}={\partial _{a}\mathbf{r}}\cdot {\partial _{b}\mathbf{r}}$:
\begin{equation}
G_{ab}=g_{ab}+\left[ \alpha g+(\alpha g)^{T}\right] _{ab}q_{3}+(\alpha
g\alpha ^{T})_{ab}q_{3}^{2}, \label{eq:metric}
\end{equation}
where
\begin{equation}
G_{a3}=G_{3a}=0,\;G_{33}=1.
\end{equation}
Here, $\alpha _{ab}$ corresponds to Weingarten curvature matrix of S \cite{dacosta,Book.holland.2013}. The equation (\ref{eq:metric}) tells us that it is possible to write the three-dimensional metric tensor as a sum of the metric tensor of the surface and the term depending on the normal direction. It is an essential feature in this approach, having an important implication: The kinetic term in the Hamiltonian is separable. In other words, we can separate the kinetic part of the hamiltonian in two contributions, the first one corresponding to internal variables (in the surface S), and the other one related to the external variable (normal direction). Explicitly, we have
\begin{equation}
\hat{H}\psi=-\frac{\hbar ^{2}}{2M_{1}}\triangle \psi -\frac{\hbar ^{2}}{2M_{2}}\left(
\frac{\partial ^{2}}{\partial q_{3}^{2}}+\frac{\partial \left( \ln \sqrt{G}
\right) }{\partial q_{3}}\frac{\partial }{\partial q_{3}}\right) \psi,
\label{laplaciano_h}
\end{equation}
The first term on the right side of (\ref{laplaciano_h}) corresponds to the kinetic term on the surface S, which $\triangle$ indicating the corresponding laplacian in the coordinates $(q_{1},q_{2})$. The second term corresponds to the laplacian for the normal coordinate. Since these terms are independent, we can admit the surface and the normal direction having different effective masses. Also, we are supposing we just have interest in the study of the dynamics of electrons on the surface S. Thus, we can imagine that the surface S corresponds to a given material with effective mass $M_1$ while the region perpendicular to S contains a different material with a different effective $M_2$.
An example of a study dealing with two different effective masses can be accessed in \cite{vinasco2018effects}. In addition, we can justify our interest in investigate anisotropic mass since some semiconductors materials, like Ge and Si, present ellipsoidal energy surfaces, related to different effective masses depending on the direction \cite{pierret1987advanced}.
Now, we need to confine the particle on the surface. In order to achieve this, a potential $
V_{\lambda }(q_{3})$ is introduced, where $\lambda$ is a parameter to measure the strength of the confinement \cite{dacosta}. This way, we have
\begin{align}
-\frac{\hbar ^{2}}{2M_{1}}\triangle \psi &-\frac{\hbar ^{2}}{2M_{2}}\left(
\frac{\partial ^{2}}{\partial q_{3}^{2}}+\frac{\partial \left( \ln \sqrt{G}
\right) }{\partial q_{3}}\frac{\partial }{\partial q_{3}}\right) \psi  \notag
\\
&+V_{\lambda }\left( q_{3}\right) \psi =i\hbar \frac{\partial \psi }{\partial t}
 \,.  \label{new}
\end{align}
The wave function can be written as
\begin{equation}
\psi (q_{1},q_{2},q_{3})=\left[ 1+\mathrm{Tr}(\alpha )q_{3}+\det (\alpha
)q_{3}^{2}\right] ^{-\frac{1}{2}}\chi (q_{1},q_{2},q_{3})\;.
\label{eq:chichi}
\end{equation}
In addition, it is possible to separate the wavefunction in the following way:
\begin{equation}
\chi (q_{1},q_{2},q_{3})=\chi_{S}(q_{1},q_{2})\chi_{n}(q_{3}),
\label{new_2}
\end{equation}
where $\chi_{S}(q_{1},q_{2})$ is the wavefunction corresponding to the surface and $\chi_{n}(q_{3})$ is the wavefunction for the normal coordinate.
Now, we will see the effect of the confining potential. When $\lambda \rightarrow \infty$, the potential confines the particle on S, in such way that we can consider $q_{3} \rightarrow 0$ in all the terms of the hamiltonian, except in the term involving the confining potential itself. Effectively, the particle is subjected to step potential barriers on both sides of S. As a result, the Schr\"{o}dinger equation gets
\begin{align}
&-\frac{\hbar ^{2}}{2M_{1}}\left[ \frac{1}{\sqrt{g}}
\partial _{a}\left( \sqrt{g}g^{ab}\partial _{b}\chi \right) \right]
\notag \\&-\frac{\hbar ^{2}}{2M_{2}}\left( \left[ \frac{1}{2}
\mathrm{Tr}(\alpha )\right] ^{2}-\det (\alpha )\right)\chi
\\&-\frac{\hbar ^{2}}{2M_{2}}\left( \partial
_{3}\right) ^{2}\chi \notag+V_{\lambda }(q_{3})\chi=i\hbar \partial _{t}\chi.
\label{H_geometric_potential}
\end{align}
This expression contains the following geometric potential
\begin{equation}
V_{S}(q_{1},q_{2})=-\frac{\hbar ^{2}}{2M_{2}}\left( \left[ \frac{1}{2}
\mathrm{Tr}(\alpha )\right] ^{2}-\det (\alpha )\right),
\label{Vs}
\end{equation}
and $g$ indicates the determinant of $g_{ab}$.
The term $\frac{1}{2}
\mathrm{Tr}(\alpha )$ in (\ref{Vs}) is the mean curvature, which can be written in terms of the principal curvatures $\kappa_{1}$ and $\kappa _{2}$:
\begin{equation}
\mathcal{M}=\frac{1}{2}Tr\left(\alpha \right)=\frac{1}{2}\left(\kappa _{1}+\kappa_{2} \right) \label{M};
\end{equation}
The term $\det \left(\alpha \right)$ corresponds to the Gaussian curvature:
\begin{equation}
\mathcal{K}_{G}=\det \left(\alpha \right)=\kappa _{1}\kappa_{2}.\label{K}
\end{equation}
These equations show how the geometric potential depends on the curvature of the surface. In addition, it is worth noting that $V_s$ depends on the effective mass $M_2$ in the normal direction. It is due to the fact that the geometric potential arises when we confine the particle on the surface S and take the limit $q_{3}\rightarrow 0$. This way, the potential $V_s$ contains information about the normal direction, in such way the mass $M_2$ affects the particle dynamics on the surface S. It is a consequence of considering a two-dimensional region immersed in a three-dimensional one. If we start considering a purely two-dimensional region, the geometric potential does not manifests, since it is not possible taking into account the influence of its neighborhood.
Now, we can reach the main goal of this section. From the discussion above, it is possible to write a Schr\"{o}dinger equation for the normal coordinate and another one for the surface:
\begin{equation*}
\hbar \partial _{t}\chi_{n}=-\frac{\hbar ^{2}}{2M_{2}}(\partial
_{3})^{2}\chi _{n}+V_{\lambda }(q_{3})\,\chi _{n},
\end{equation*}
\begin{align}
\hbar \partial_{t}\chi_{S}=&-\frac{\hbar ^{2}}{2M_{1}}\left[ \frac{1}{\sqrt{
g}}\partial _{a}\left( \sqrt{g}g^{ab}\partial _{b}\chi_{S}\right) \right] \notag \\&-
\frac{\hbar ^{2}}{2M_{2}}\left[\frac{1}{4}\mathcal{M}^{2}-
\mathcal{K}_{G}\right] \chi_{S},  \label{eq:shrodecop}
\end{align}
where the indexes $a,b=1,2$ corresponds to the surface S.
We have considered a metric tensor such that
\begin{equation}
G_{ij}=
\begin{cases}
g_{ab} \hspace{0.2cm} \mbox{if} \hspace{0.2cm} i,j=a,b=1,2 ; \\
G_{33}=1; \\
G_{i3}=G_{3j} \hspace{0.2cm} \forall \hspace{0.2cm} i,j=1,2.
\end{cases}
\end{equation}

\section{Quantum harmonic oscillator on a helicoid}
In this section, we consider the problem of a quantum particle constrained to a helicoidal surface. We can use the following equations to parametrize a helicoid \cite{geometry}:
\begin{equation}
\begin{cases}
x=\rho \cos (\omega z), \\
y=\rho \sin (\omega z), \\
z=z,
\end{cases}
\label{chl}
\end{equation}
with $\omega =2\pi S$. The number of complete twists per unit length of the helicoid is given by $S$. $\rho $ measures the radial
distance from the $z$-axis.
The corresponding metric tensor is
\begin{equation}
g_{ab}=\left(
\begin{array}{ccc}
1 & 0  \\
0 & 1+\omega ^{2}\rho ^{2}
\end{array}
\right),
\end{equation}
and the infinitesimal line element is
\begin{equation}
ds^{2}=d\rho ^{2}+(1+\omega ^{2}\rho ^{2})\,dz^{2}\;.
\label{E:lineelementheli}
\end{equation}
\begin{figure}[!h]
\centering	
\includegraphics[width=0.4\textwidth]{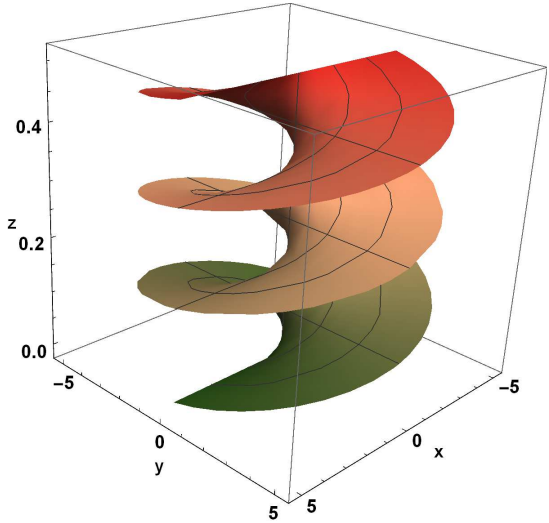}
\caption{Geometry of the helicoid represented by the line element (\ref{E:lineelementheli}).}
\label{helicoidpng}
\end{figure}
Figure \ref{helicoidpng} shows a helicoid. The geometric potential in this case is
\begin{equation}
V_{S}=-\frac{\hbar ^{2}}{2M_{2}}\left( \mathcal{M}^{2}-\mathcal{K}_{G}\right)
=-\frac{\hbar ^{2}}{2M_{2}}\frac{\omega ^{2}}{(1+\omega ^{2}\rho ^{2})^{2}}
\;,  \label{E:curvpotentialheli}
\end{equation}
since the principal curvatures are given by
\begin{equation}
\kappa _{1}=\frac{\omega }{1+\omega ^{2}\rho ^{2}}, \hspace{0.4cm} \kappa _{2}=-\kappa _{1}.
\label{E:metricdeterminantheli}
\end{equation}
Our goal consists in considering an harmonic oscillator in a helicoid. This way, the particle will be subjected to an effective potential composed of a geometrical potential for the helicoidal surface and also to an harmonic oscillator potential. Let us construct the hamiltonian.  We start by considering the hamiltonian for a particle on a helicoid in the context of anisotropic mass.
The corresponding Schr\"{o}dinger in the coordinates $\rho$ and $z$ is
\begin{align}
i\hbar \partial _{t}\chi _{S}& =-\frac{\hbar ^{2}}{2M_{1}}\left[ \frac{1}{a}
\left( \partial _{z}(\frac{1}{a}\partial _{z}\chi _{S})+\partial _{\rho
}(a\partial _{\rho }\chi _{S})\right) \right]  \notag \\
& -\frac{\hbar ^{2}}{2M_{2}}\frac{\omega ^{2}}{(1+\omega ^{2}\rho ^{2})^{2}}
\chi _{s}\;,  \label{E:hamiltoniancurv}
\end{align}
with $a\equiv \sqrt{1+\omega ^{2}\rho ^{2}}$.
Following \cite{souza2018curved}, we make the separation of the variables $\chi _{S}=\exp \left( im\omega z\right) f\left( \rho \right)$, where $m\in \mathbb{N}$. In addition, the wavefunction normalization demands the transformation \cite{atanasov} $\chi
_{S}\rightarrow \frac{1}{\sqrt{a}}\chi _{S}$ in Eq. (\ref{E:hamiltoniancurv}
).
\begin{figure*}[!t]
\centering
\includegraphics[scale=0.75]{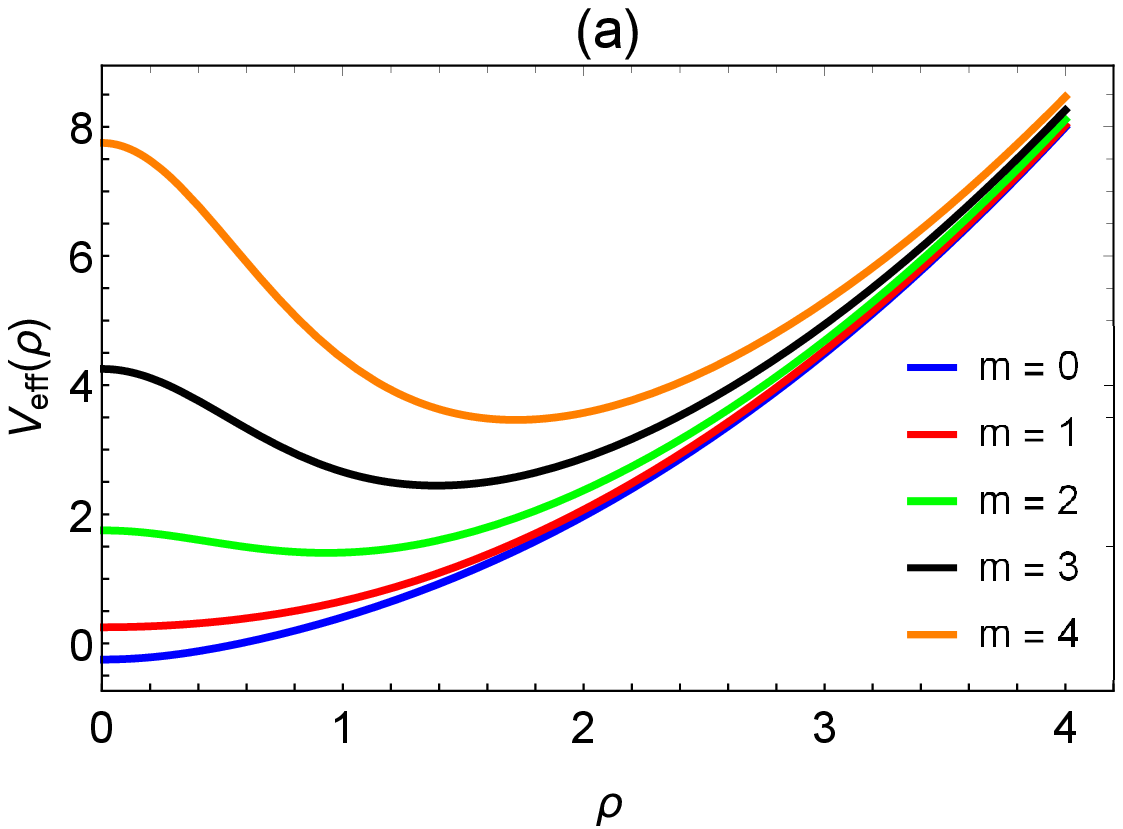}\qquad
\includegraphics[scale=0.75]{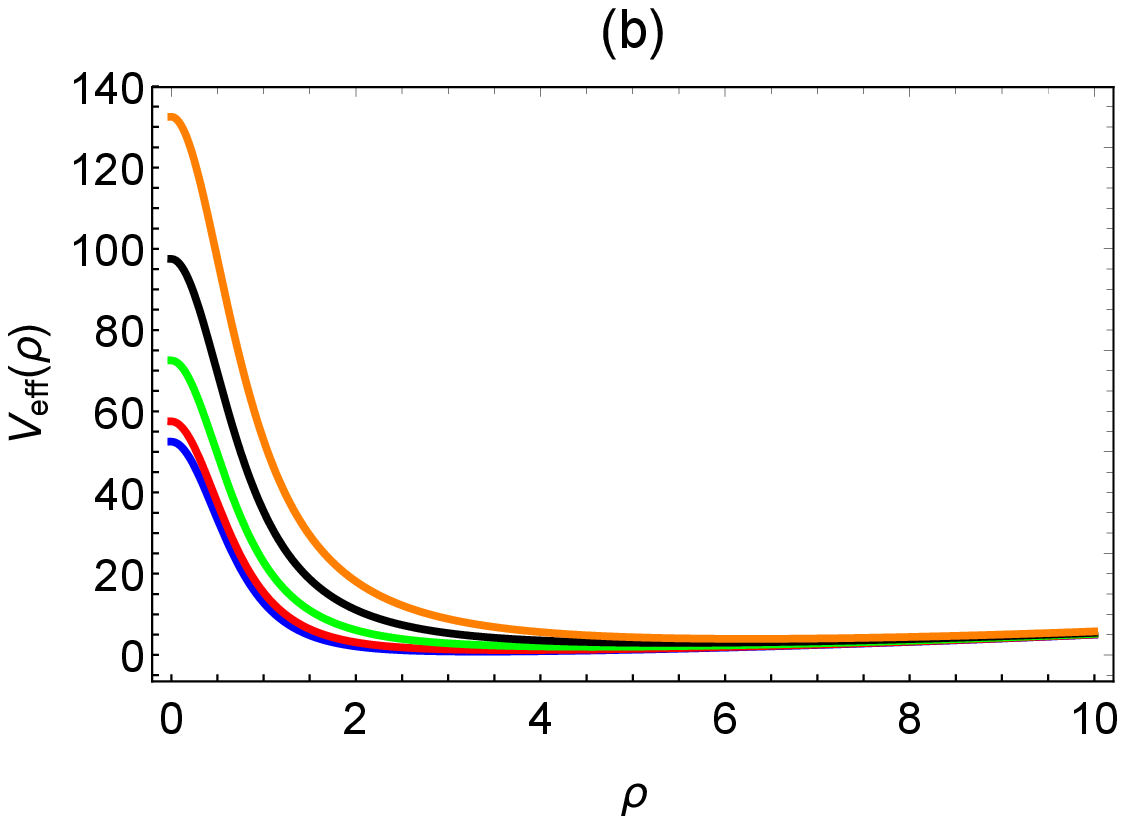}\\[07.pt]
\includegraphics[scale=0.75]{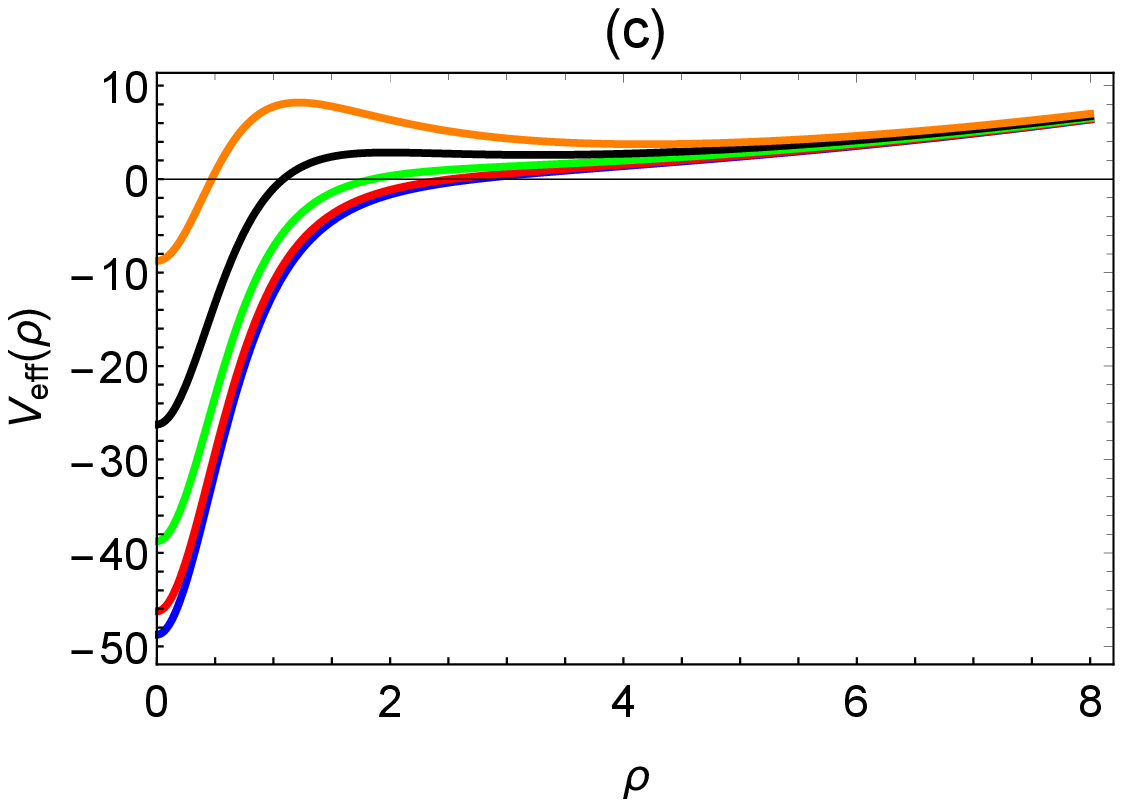}\qquad
\includegraphics[scale=0.75]{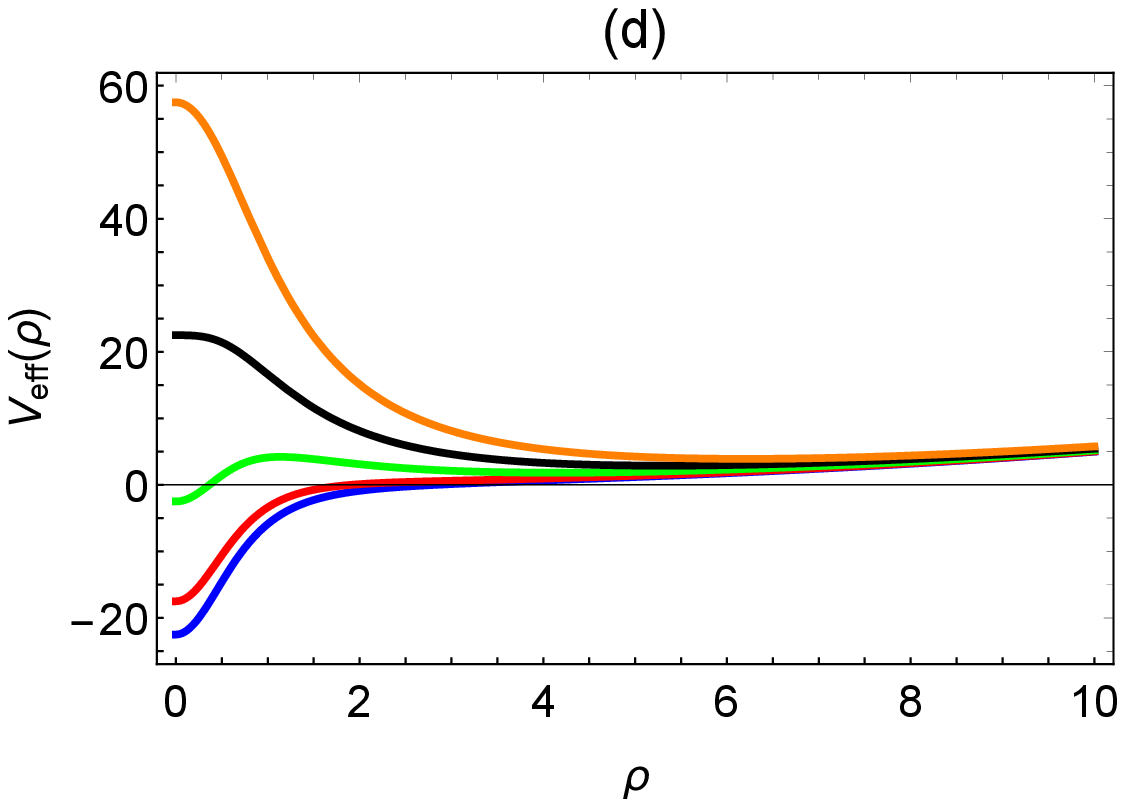}
\caption{ (Color online) The behavior of the effective potential (Eq. (
\protect\ref{Ueff2})) as a function of $\rho$. In panel (a) we consider $M_{1}=M_{2}=1$. In (b), $M_{1}=0.1$ and $M_{2}=-0.01$. In (c), $M_{1}=0.2$ and $M_{2}=0.01$. In (d), $M_{1}=0.1$ and $M_{2}=0.02$. We use  $\hbar =1$, $\protect\omega =1$ and $\Omega
=1$.} \label{Fig_2D_Eff_Pot}
\end{figure*}
After these steps, we obtain the Schr\"{o}dinger equation in the helicoidal geometry:
\begin{align}
H_{curv}\chi _{s}=&-\frac{\hbar ^{2}}{2M_{1}}\frac{d^{2}\chi _{s}}{d\rho
^{2}}+\frac{\hbar ^{2}}{2M_{1}}\Bigg[\frac{m^{2}\omega ^{2}}{1+\omega
^{2}\rho ^{2}}  \notag \\
& -\frac{\omega ^{2}}{2(1+\omega ^{2}\rho ^{2})^{2}}\left( \frac{\omega
^{2}\rho ^{2}}{2}+\frac{2M_{1}}{M_{2}}-1\right) \Bigg]\,\chi _{s}\;.
\label{E:hamiltoniancurvnorm}
\end{align}
In particular, if $M_{1}=M_{2}=m^{\ast }>0$ and $\rho =\xi $, we get the potential investigated in Ref. \cite{atanasov}. We already have the Hamiltonian for a particle in a helicoid in the context of anisotropic mass. The next ingredient we need is to include the harmonic oscillator potential to construct an effective potential and write the corresponding differential equation. In Cartesian coordinates, the harmonic potential is given by
\begin{equation}
V(x,y)=\frac{1}{4}M_{1}\Omega _{x}^{2}x^{2}+\frac{1}{4}M_{1}\Omega _{y}^{2}y^{2},
\label{vosc}
\end{equation}
where the factor $\frac{1}{4}$ was introduced for convenience.

\begin{figure}[!h]
\centering
\fbox{\begin{minipage}{3.3in}
\centering
\includegraphics[totalheight=1.38in]{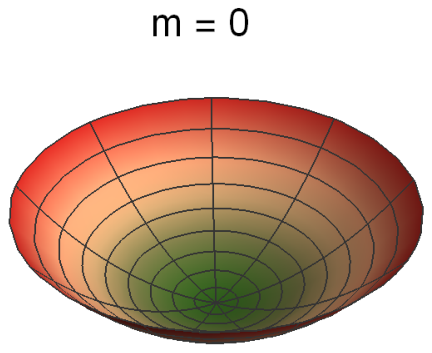}
\includegraphics[totalheight=1.38in]{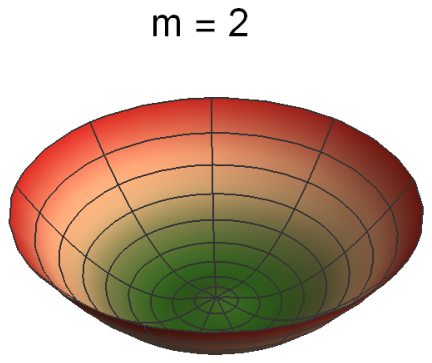}
\includegraphics[totalheight=1.38in]{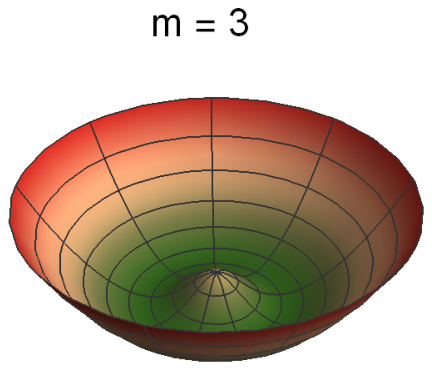}
\includegraphics[totalheight=1.38in]{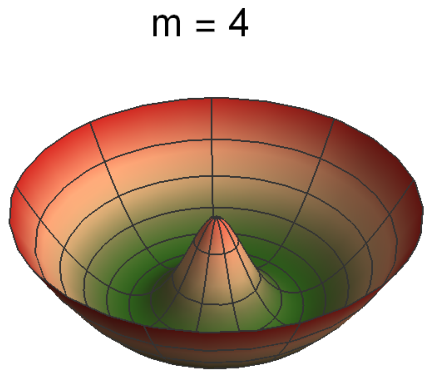}\vspace{-1.0cm}
\caption{ (Color online) A $3D$ visualization of the effective potential
sketched in Fig. \ref{Fig_2D_Eff_Pot}-(a).}
\label{Fig_Potential_3D_A}
\end{minipage}}
\end{figure}

\begin{figure}[!h]
\centering
\fbox{\begin{minipage}{3.3in}
\centering
\includegraphics[totalheight=1.38in]{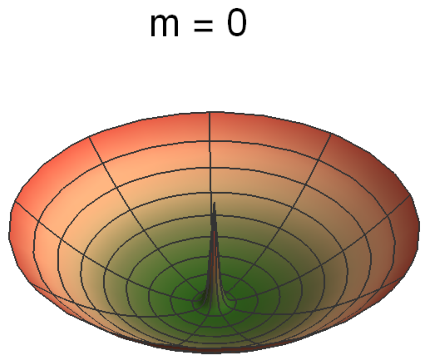}
\includegraphics[totalheight=1.38in]{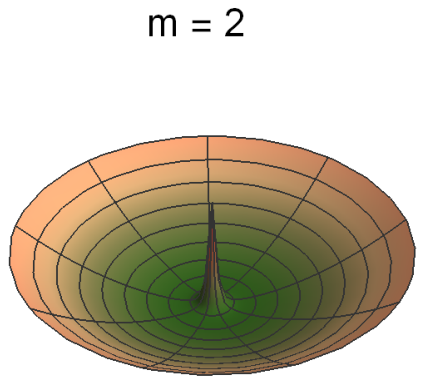}
\includegraphics[totalheight=1.38in]{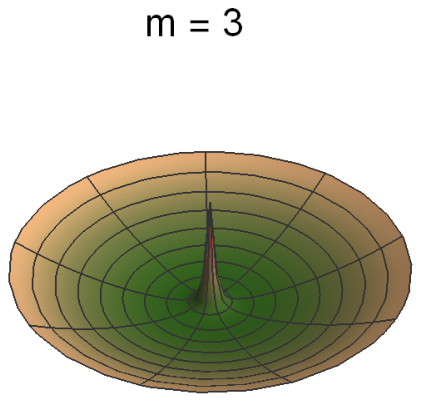}
\includegraphics[totalheight=1.38in]{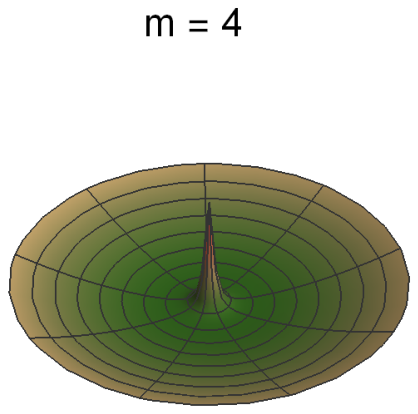}\vspace{-1.0cm}
\caption{ (Color online) A $3D$ visualization of the effective potential sketched in Fig. \ref{Fig_2D_Eff_Pot}-(b).}
\label{Fig_Potential_3D_B}
\end{minipage}}
\end{figure}

\begin{figure}[!t]
\centering
\fbox{\begin{minipage}{3.3in}
\centering
\includegraphics[totalheight=1.01in]{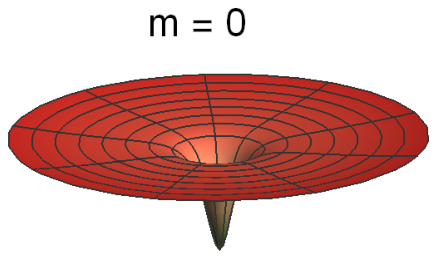}
\includegraphics[totalheight=1.01in]{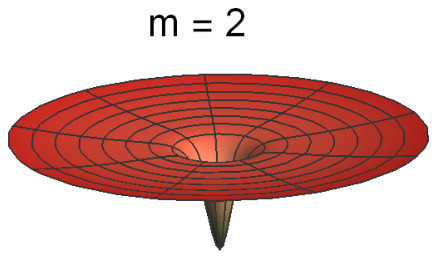}
\includegraphics[totalheight=1.01in]{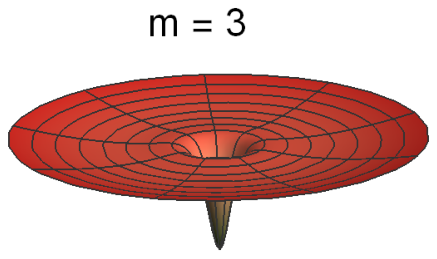}
\includegraphics[totalheight=1.01in]{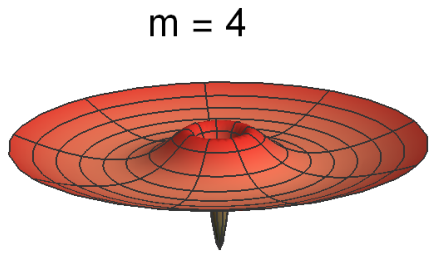}\vspace{-1.0cm}
\caption{(Color online) A $3D$ visualization of the effective potential sketched in Fig. \ref{Fig_2D_Eff_Pot}-(c).}
\label{Fig_Potential_3D_C}
\end{minipage}}
\end{figure}

\begin{figure}[!t]
\centering
\fbox{\begin{minipage}{3.3in}
\centering
\includegraphics[totalheight=0.95in]{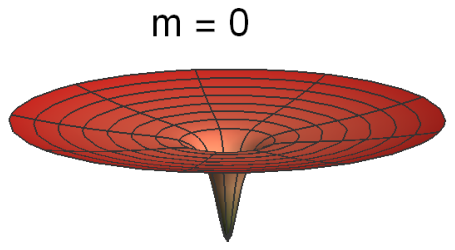}
\includegraphics[totalheight=0.95in]{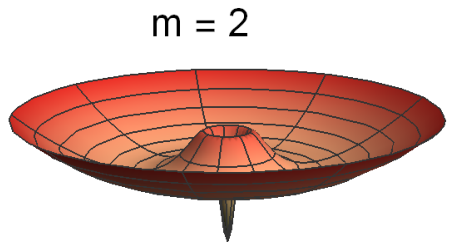}
\includegraphics[totalheight=0.95in]{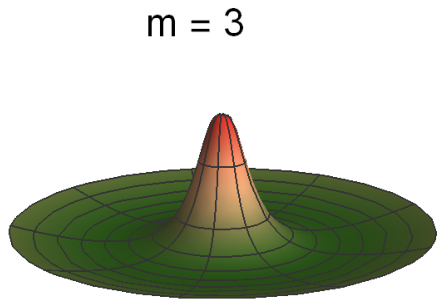}
\includegraphics[totalheight=0.95in]{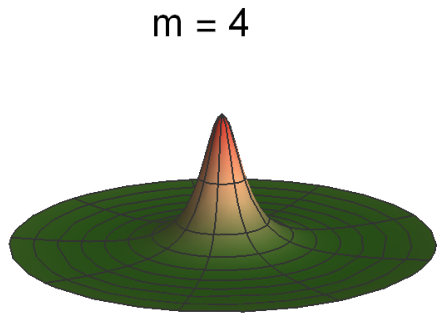}\vspace{-0.3cm}
\caption{ (Color online) A $3D$ visualization of the effective potential sketched in Fig. \ref{Fig_2D_Eff_Pot}-(d).}
\label{Fig_Potential_3D_D}
\end{minipage}}
\end{figure}

 We are interested in the motion of an harmonic oscillator with a single frequency $
\Omega _{x}=\Omega _{y}=\Omega >0$, so that the potential (\ref{vosc})
written in the coordinates (\ref{chl}) reads as
\begin{equation}
V(\rho )=\frac{1}{2}M_{1}\Omega ^{2}\rho ^{2},  \label{oscillator}
\end{equation}
which can be included into the Eq. (\ref{E:hamiltoniancurv}) by means of the
substitution $E\rightarrow E-V\left( \rho \right) $, resulting in the radial equation
\begin{equation}
-\frac{\hbar ^{2}}{2M_{1}}\frac{d^{2}f\left( \rho \right) }{d\rho ^{2}}
+V_{eff}\left( \rho \right) f\left( \rho \right) =Ef\left( \rho \right) ,
\label{Schreff}
\end{equation}
where
\begin{align}
V_{eff}\left( \rho \right) =\frac{\hbar ^{2}}{2M_{1}}&\Bigg[ \frac{\omega
^{2}m^{2}}{1+\omega ^{2}\rho ^{2}}-\frac{1}{2}\frac{\omega ^{2}}{\left(
\omega ^{2}\rho ^{2}+1\right) ^{2}}\notag \\ & \times \left( \frac{1}{2}\omega ^{2}\rho ^{2}+
\frac{2M_{1}}{M_{2}}-1\right) \Bigg] +\frac{1}{2}M_{1}\Omega ^{2}\rho ^{2}
\label{Ueff2}
\end{align}
is the effective potential. Figure \ref{Fig_2D_Eff_Pot} shows plots of $V_{eff}$, taking some specific values for the effective masses. We consider $M_1$ always being positive, while $M_2$ can be either positive or negative. When $M_{1}=M_{2}=1$, the resulting potential is parabolic for $m=0$ and $m=1$ while for $m=2$, $3$ and $4$ it exhibits a narrow binding region (Fig. \ref{Fig_2D_Eff_Pot}-(a)). These two characteristics have important applications in mesoscopic physics. For example, the former can be considered as a model of a quantum dot while the last one describes a narrow ring. The appearance of these localization regions is a consequence of the helicoidal geometry. However, this characteristic is also manifested when $M_{1} \neq M_{2}$. Figures \ref{Fig_2D_Eff_Pot}-(b)-(d) clearly show that the presence of anisotropic mass reveals new exotic characteristics for the effective potential. For example, in the particular case when $M_{1}=0.1$ and $M_{2}=-0.01$, a localized wide binding region appears (Fig. \ref{Fig_2D_Eff_Pot}-(b)). This situation can be interpreted as the analog of a mesoscopic wide ring. Depending on the given anisotropy, more than one binding region may appear. This is exemplified when we have $M_{1}=0.2$ and $M_{2}=0.01$, which has two binding regions (Fig. \ref{Fig_2D_Eff_Pot}-(c)). The first located region (near the origin) describes a potential well and the second a wide ring. This model also presents mixed characteristics, such as those present in previous cases. An example of this occurs when $M_{1}=0.1$ and $M_{2}=0.02$, where we can observe the potential profile present in Figs. \ref{Fig_2D_Eff_Pot}-(b) and \ref{Fig_2D_Eff_Pot}-(c). Figures \ref{Fig_Potential_3D_A}-\ref{Fig_Potential_3D_D} show the $3D$ profiles of Fig. \ref{Fig_2D_Eff_Pot}, with the exception of the case with $m=1$, which has an equivalent shape to the curve with $m=0$ in the four cases. The symmetry of the quantum number $m$ in the effective potential allows us to visualize more clearly the regions that allow bound states in the $3D$ figures.

Particularly, the case in which $M_2 < 0$ consists of a quantum mechanical analog of a hyperbolic metamaterial, as discussed in \cite{souza2018curved}. Metamaterials \cite{Smith788} are quite interesting because of their unique possibilities of investigations involving negative refractive indexes and technological developments \cite{liu2015metamaterials}. Also, metamaterials allow the emulation of theoretical cosmology scenarios in the context of electrodynamics \cite{PhysRevD.96.105012}.
Thus, basically, it is possible to analyze the harmonic oscillator taking into account three different possibilities: i) an isotropic sample, where the helicoid  is surrounded by the same material of the helicoidal surface, ii) an anisotropic material, and iii) an electronic analogue of a hyperbolic material.
The effective potential is an even function. This way, the Hamiltonian commutes with the Parity Operator. Then, these operators can share a mutual basis of eigenstates. As a consequence, we expect that the solution can accommodate either symmetric or antisymmetric solutions, like in the case of the commun quantum harmonic oscillator.

\begin{figure*}[!t]
\centering
\includegraphics[scale=0.55]{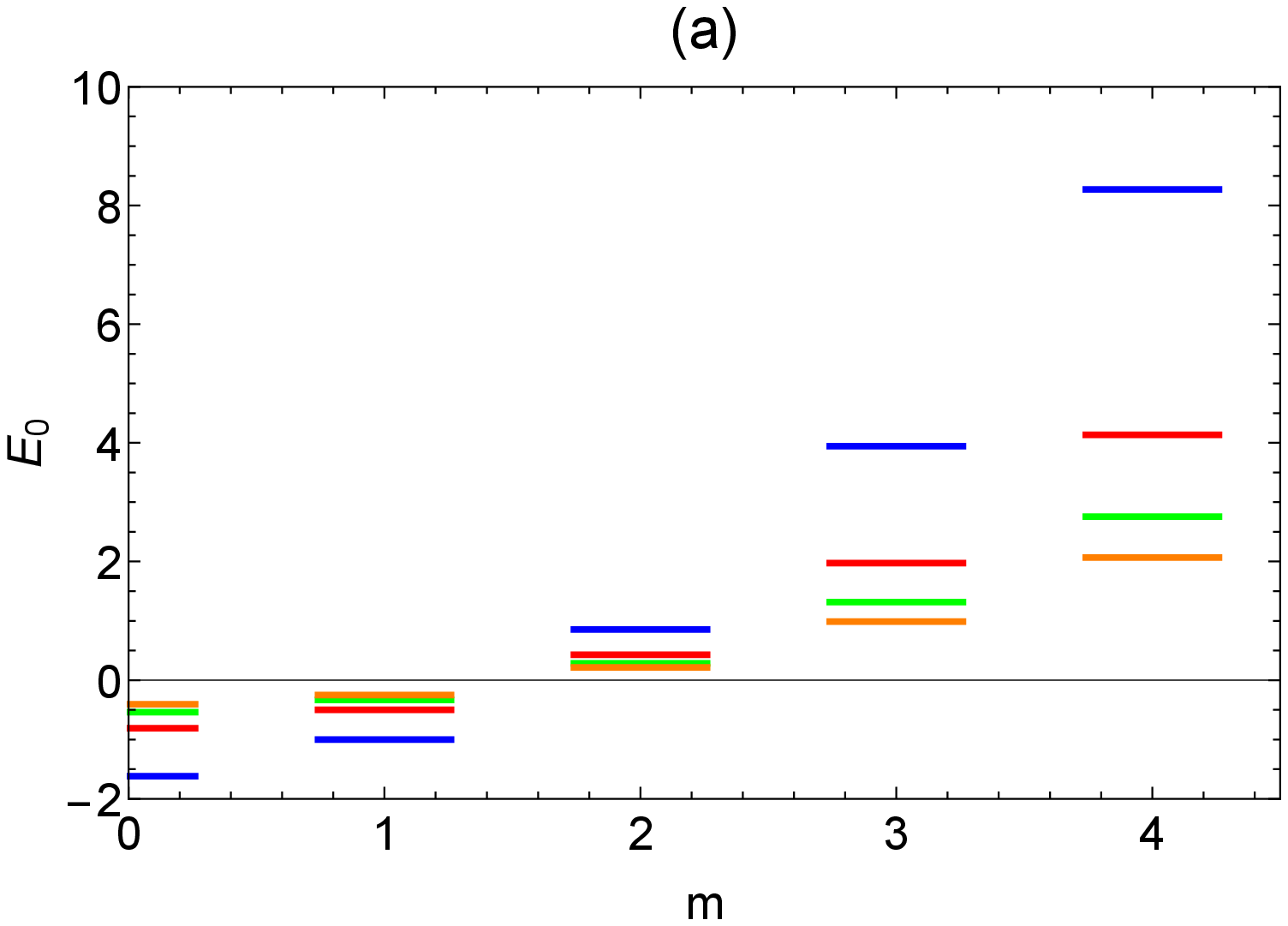} \qquad
\includegraphics[scale=0.55]{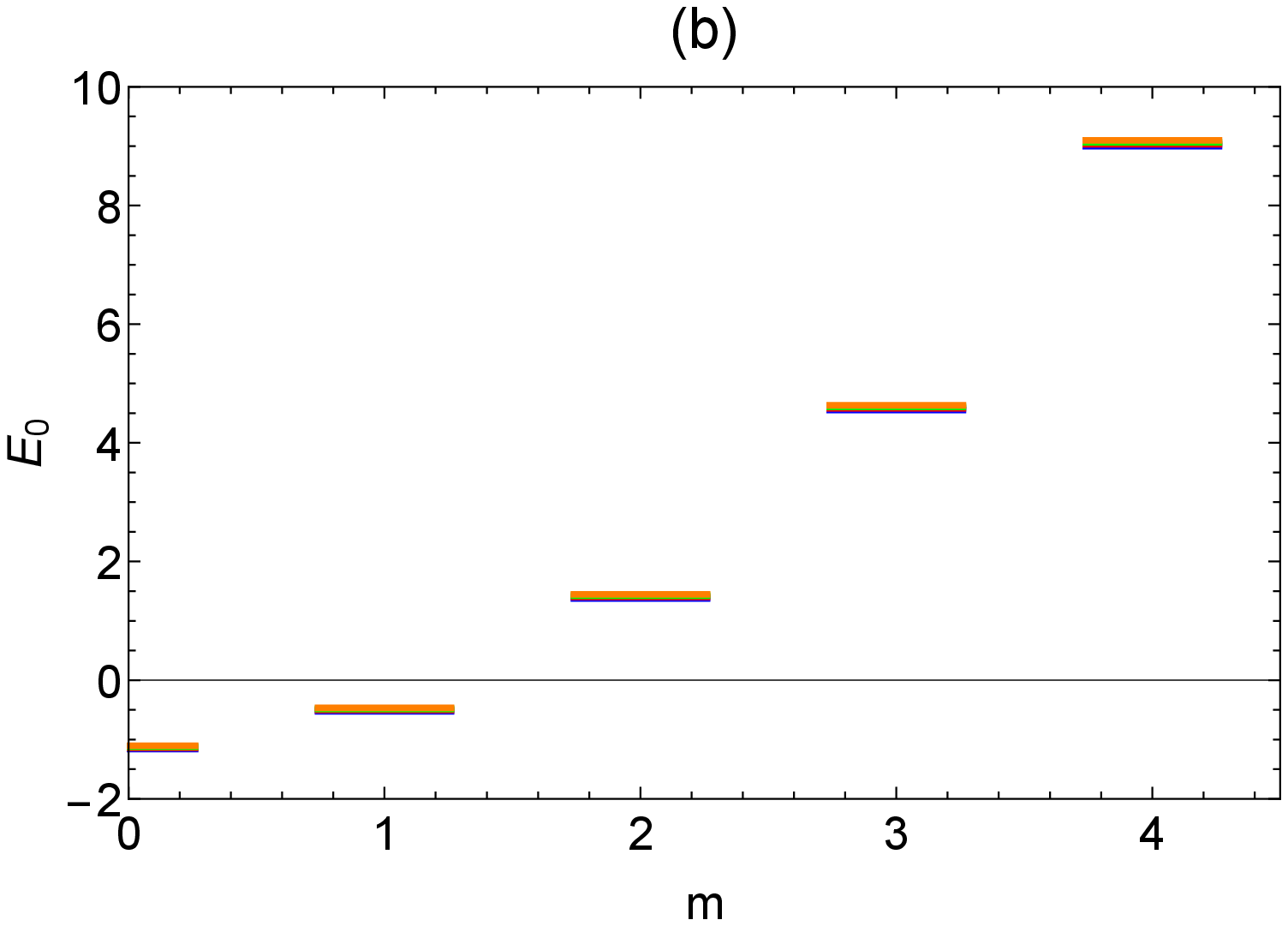}\\[7pt]
\includegraphics[scale=0.55]{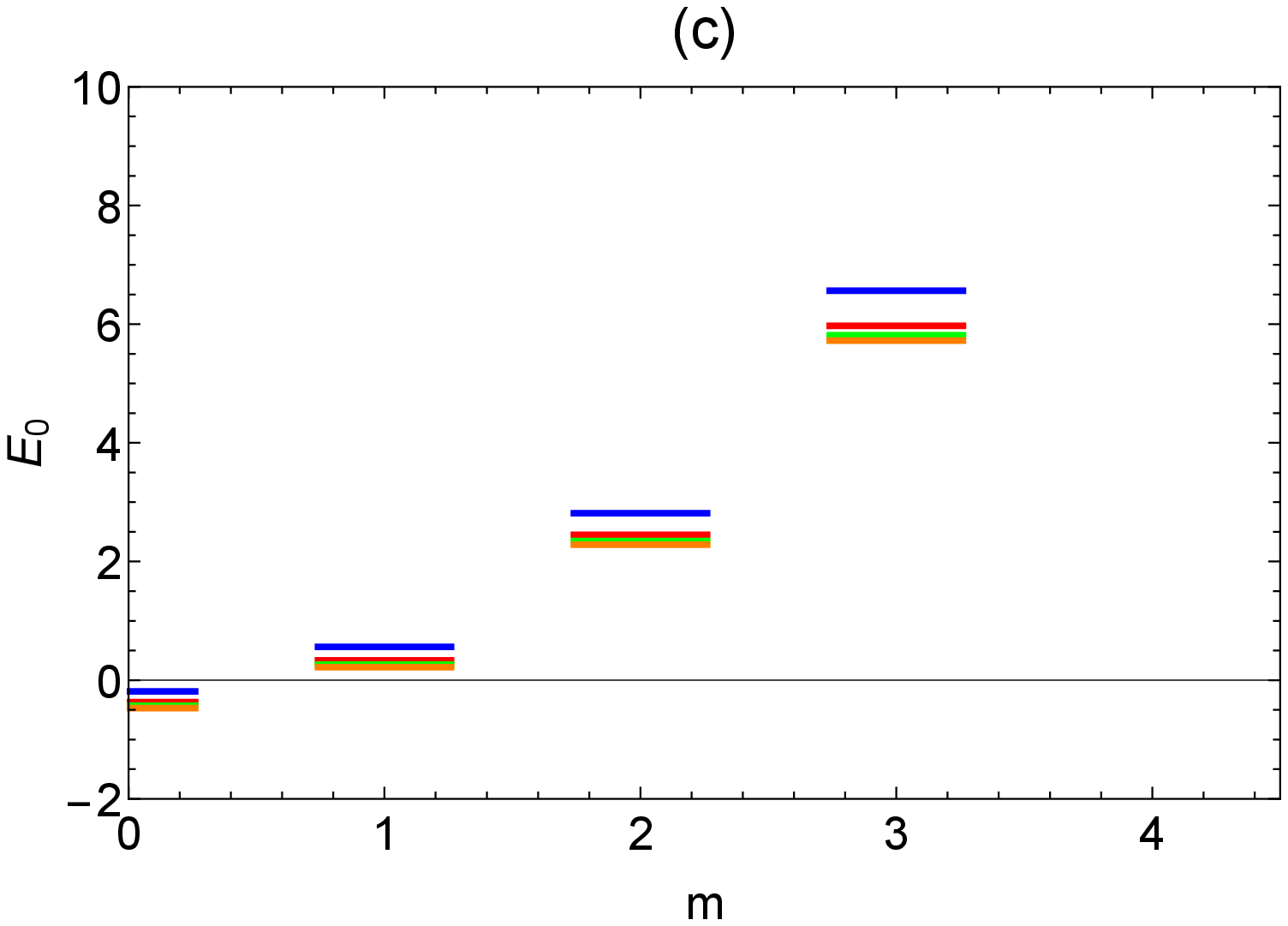} \qquad
\includegraphics[scale=0.55]{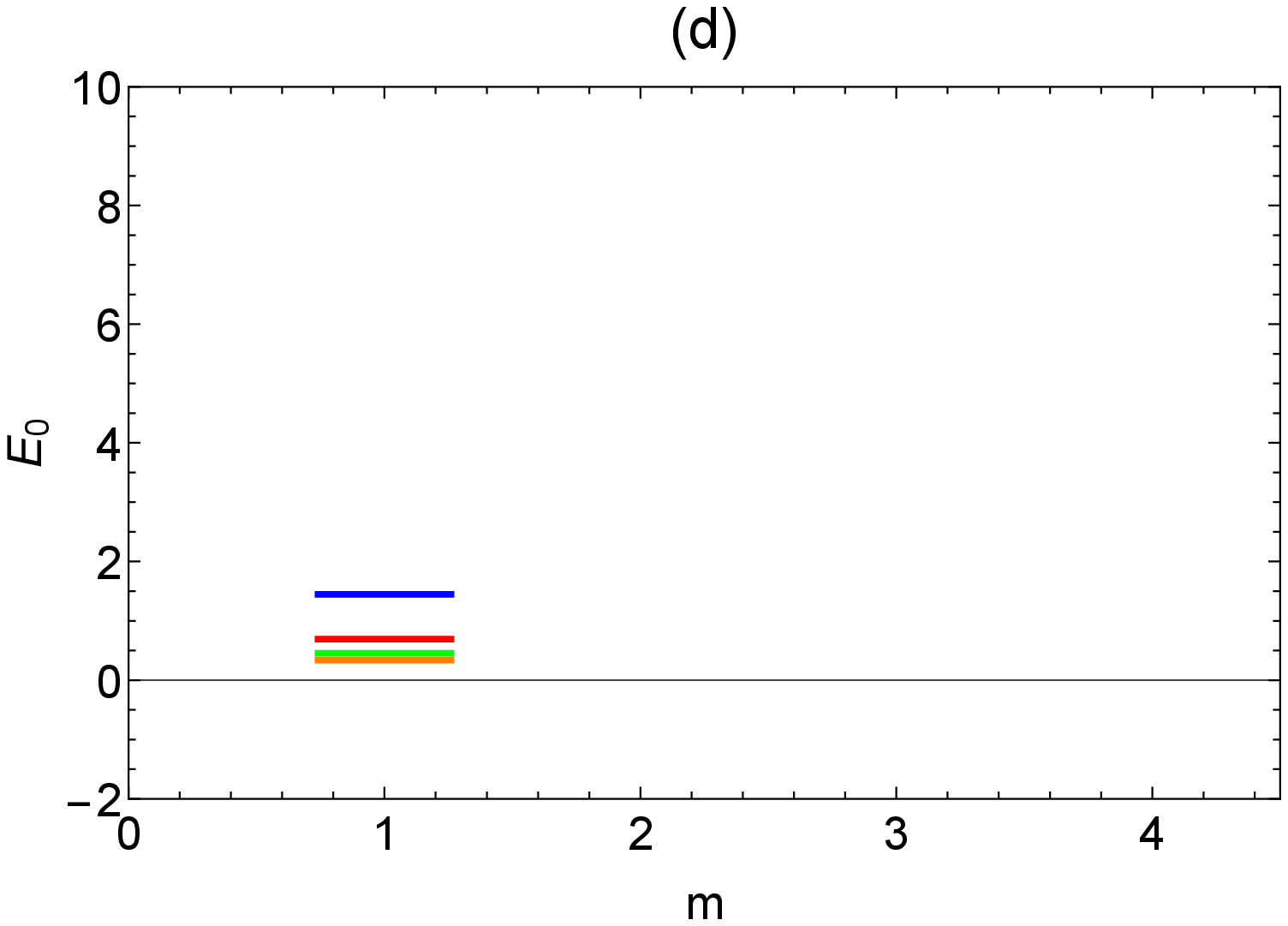}
\caption{ (Color online) Sketch of the energy levels $E_{0m}$ as a function of the quantum number $m$. We use $\hbar =1$, $\protect\omega =1$ and $\Omega =1$.}\label{Energy_Ground_State}
\end{figure*}

\begin{figure*}[!t]
\centering
\includegraphics[scale=0.87]{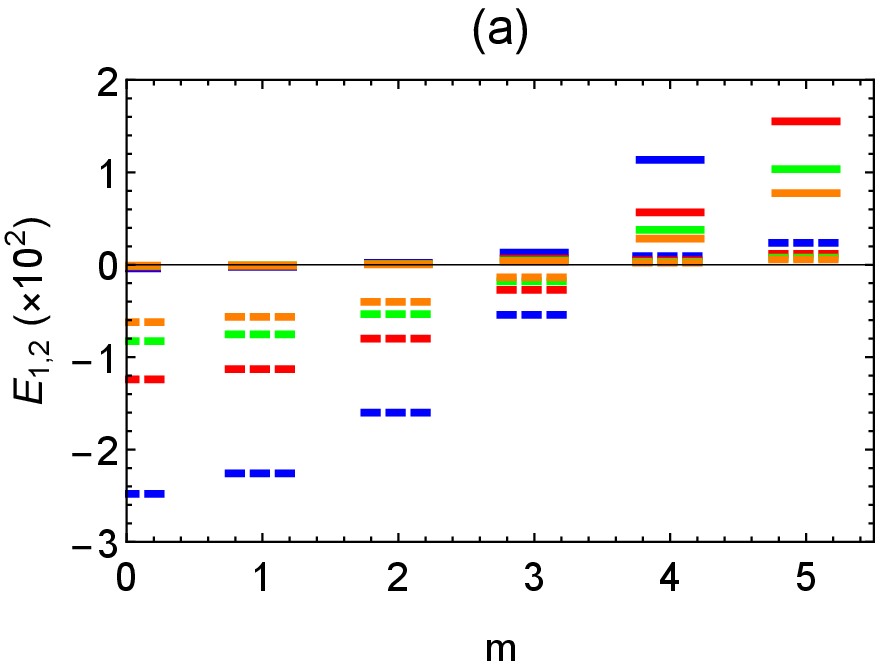}\qquad
\includegraphics[scale=0.87]{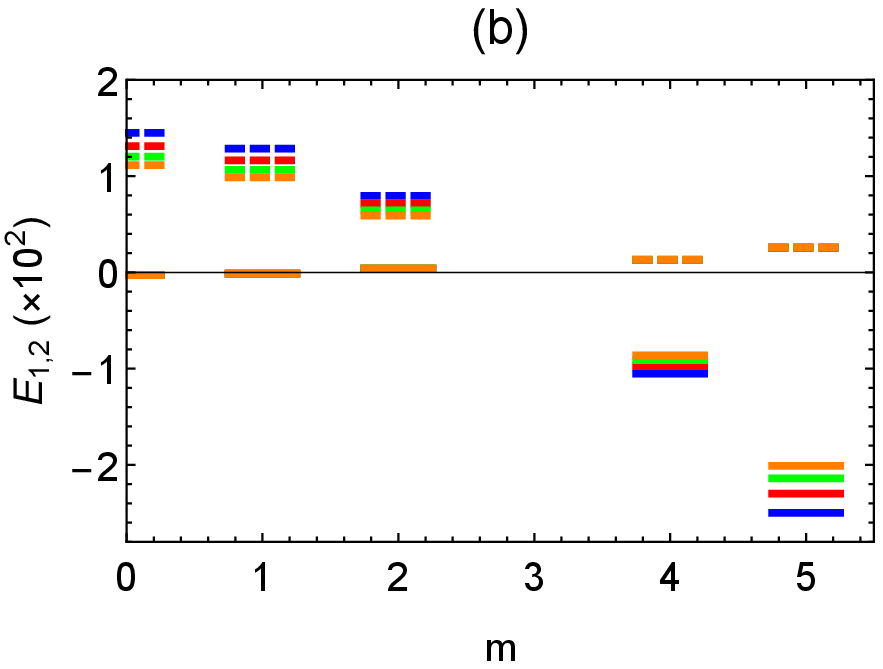}\\[7pt]
\includegraphics[scale=0.89]{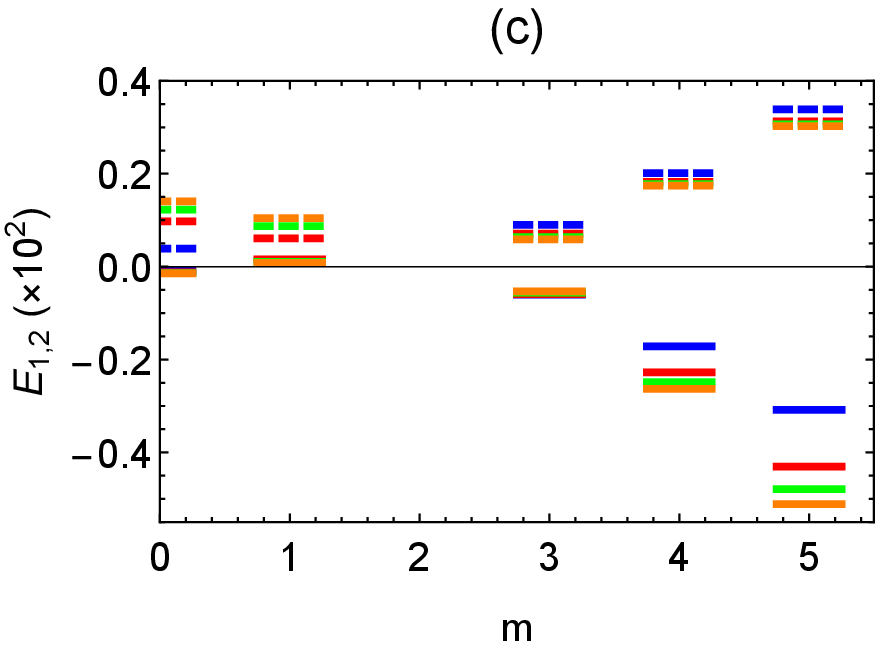}\qquad
\includegraphics[scale=0.87]{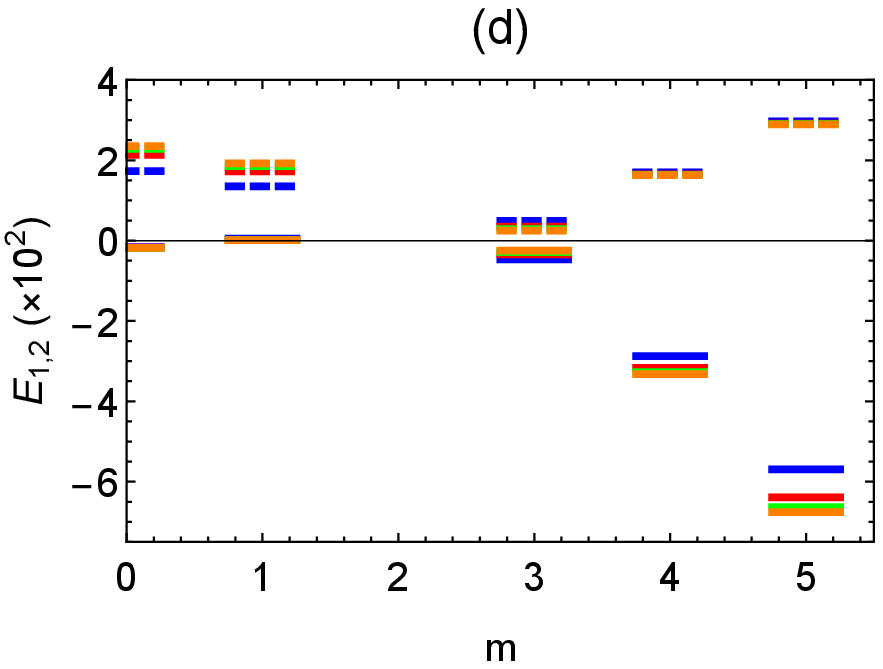}
\caption{ (Color online) Sketch of the energies with $n=1$ considering some specific values of $M_{1}$ and $M_{2}$. In panel (a) we consider $(M_{1},M_{2})=\{(1,1),(2,2),(3,3),(4,4)\}$. In (b), $(M_{1},M_{2})=\{(1,2.1),(1,2.2),(1,2.3),(1,2.4)\}$. In (c), $(M_{1},M_{2})=\{(1,-4),(1,-5),(1,-6),(1,-7)\}$. In (d), $(M_{1},M_{2})=\{(0.1,-1),(0.1,-2),(0.1,-3),(0.1,-4)\}$. The solid lines correspond to energies $E_{1}$ (Eq. (\ref{Om1E1})) while dotted lines correspond to energies $E_{2}$ (Eq. (\ref{Om1E2})). We use  $\hbar =1$, $\protect\omega =1$ and $\Omega
=1$.}
\label{Fig_2D_Energy_n1}
\end{figure*}

Equation (\ref{Schreff}) can be written as
\begin{align}
&\frac{d^{2}f\left( \rho \right) }{d\rho ^{2}}+\left[\frac{M_{1}}{M_{2}}\frac{\omega ^{2}}{\left( 1+\omega ^{2}\rho
^{2}\right) ^{2}}-\varpi ^{2}\rho ^{2}+k^{2}\right]\,f\left( \rho \right)\notag\\&+\left[ \frac{3\omega ^{4}\rho
^{2}}{4\left( 1+\omega ^{2}\rho ^{2}\right) ^{2}}-\frac{\omega ^{2}}{2\left(
1+\omega ^{2}\rho ^{2}\right) }-\frac{\omega ^{2}m^{2}}{1+\omega ^{2}\rho
^{2}}\right] f\left( \rho \right) =0,  \label{HeunCA}
\end{align}
where $k^{2}=2M_{1}E/\hbar ^{2}$, $\varpi ^{2}=M_{1}^{2}\Omega ^{2}/\hbar
^{2}$. Equation (\ref{HeunCA}) is of the Heun's confluent differential
equation type \cite{Book.Ronveaux.1995,Book.Slavianov.2000}
\begin{align}
{\Phi }^{\prime \prime }\left( z\right) +&\left( \alpha +\frac{\beta +1}{z}+
\frac{\gamma +1}{z-1}\right) {\Phi }^{\prime }(z)\notag \\&-\frac{1}{2}\,\left( \frac{
\mu }{z}+\frac{\nu }{z-1}\right) {\Phi }(z)=0  \label{HeunC}
\end{align}
with
\begin{align}
\mu  &=\frac{1}{2}\left( \alpha -\beta -\gamma +\alpha \beta -\beta \gamma
\right) -\eta , \label{mu}\\
\nu  &=\frac{1}{2}\left( \alpha +\beta +\gamma +\alpha \gamma +\beta \gamma
\right) +\delta +\eta ,\label{nu}
\end{align}
The solution to Eq. (\ref{HeunC}) is computed as a power series expansion around the origin $z=0$, a regular singular point with a radius of convergence $\left\vert z\right\vert <1$, given by
\begin{equation}
{\Phi }(z)=\sum\limits_{s=0}^{\infty }\upsilon _{s}\left( \alpha ,\beta
,\gamma ,\delta ,\eta ,z\right) z^{s}=\mathit{HeunC}\left( \alpha ,\beta
,\gamma ,\delta ,\eta ,z\right),   \label{series}
\end{equation}
where the coefficients $\upsilon _{s}$ satisfy a three-term recurrence relation
\begin{equation}
A_{s}\upsilon _{s}=B_{s}\upsilon _{s-1}+C_{s}\upsilon _{s-2},  \label{recorr}
\end{equation}
with initial conditions
\begin{eqnarray}
\upsilon _{-1} &=&0, \\
\upsilon _{0} &=&1,
\end{eqnarray}
where
\begin{align}
A_{s} &=1+\frac{\beta }{s}, \\
B_{s} &=1+\frac{1}{s}\left( \beta +\gamma -\alpha -1\right) \notag \\&+\frac{1}{n^{2}}
\left[ \eta -\frac{1}{2}\left( \beta +\gamma -\alpha \right) -\frac{\alpha
\beta }{2}+\frac{\beta \gamma }{2}\right] , \\
C_{s} &=\frac{\alpha }{s^{2}}\left( \frac{\delta }{\alpha }+\frac{\beta
+\gamma }{2}+n-1\right).
\end{align}
\begin{table}[!h]
     \begin{tabular}{p{8cm}}
     \centering
     \includegraphics[width=0.46\textwidth, height=2cm]{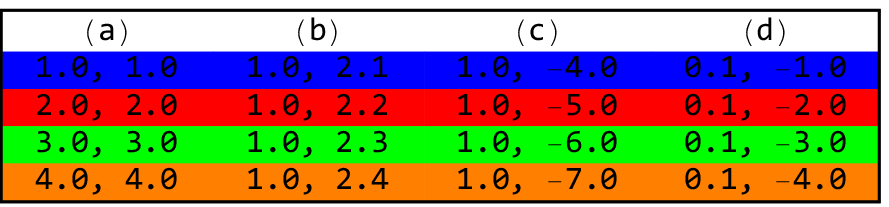}
      \end{tabular}
      \caption{Values of ($M_{1}$, $M_{2}$) used in the sketches in Figures \ref{Energy_Ground_State} and \ref{Fig_2D_Energy_n1}.}
      \label{Table1}
      \end{table}
      In Table \ref{Table1}, the columns (a)-(d) refer to Figs. \ref{Energy_Ground_State}-\ref{Fig_2D_Energy_n1}\,(a)-(d), respectively. In the order in which the columns are presented, each value pair corresponds to ($M_{1}$, $M_{2}$). The colors blue, red, green and orange refer to the energy levels as a function of $m$, following the order in which they appear in the figures.
This confluent Heun function $\mathit{HeunC}\left( \alpha ,\beta
,\gamma ,\delta ,\eta ,z\right)$ must reduce to
a polynomial, since otherwise it would increase exponentially as $\rho
\rightarrow \infty $. To reduce a confluent Heun function to a confluent
Heun polynomial of degree $n$ we need two successive terms in the three-term
recurrence relation, Eq. (\ref{recorr}), to vanish, halting the infiniteserie s, Eq. (\ref{series}). This requirement results in two termination conditions, both needed to be satisfied simultaneously \cite{PRB.2016.94.165407,JMP.2013.54.072101,PRD.2009.80.124001,JPA.2009.43.035203}
\begin{align}
& \frac{\delta }{\alpha }+\frac{\beta +\gamma }{2}+n+1=0,\;\;n=1,2,\ldots
\label{cndA} \\
& \Delta _{n+1}\left( \mu \right) =0.  \label{cndB}
\end{align}
The solution to Eq. (\ref{HeunCA}) is given by
\begin{align}
f\left( \rho \right)  =&\mathit{c}_{m}\,\left( {\omega }^{2}{\rho }
^{2}+1\right) ^{\frac{\,1}{2}\left( \gamma +1\right) }e{^{-\frac{1}{2}\varpi
{\rho }^{2}}}\notag \\ &\times \mathit{HeunC}\left( \alpha ,\beta ,\gamma ,\delta ,\eta ,-{
\omega }^{2}{\rho }^{2}\right)   \notag \\
& +\mathit{d}_{m}\,\left( {\omega }^{2}{\rho }^{2}+1\right) ^{\frac{\,1}{2}
\left( \gamma +1\right) }\rho \,e{^{-\frac{1}{2}\varpi {\rho }^{2}}}\notag \\ &\times\mathit{
HeunC}\left( \alpha ,-\beta ,\gamma ,\delta ,\eta ,-{\omega }^{2}{\rho }
^{2}\right),
\label{solutionHC}
\end{align}
where
\begin{align}
\alpha  &=\frac{{\varpi }}{{{\omega }^{2}}},\;\;\beta =-\frac{1}{2},\;\;\gamma =\frac{x}{2},\;\;\delta =-\frac{{{k
}^{2}}}{{4{\omega }^{2}}},\;\; \notag\\ x&=\sqrt{\frac{4M_{{1}}}{M_{{2}}}+1},\;\; \eta=\frac{1}{8}\left( 3-2\,\;{m}^{2}\right) +\frac{M_{{1}}}{4M_{{2}}}+
\frac{{k}^{2}}{4{\omega }^{2}},\;\;\notag \\
\varpi &=\frac{M_{1}\Omega }{\hbar },\;\;k^{2}=
\frac{2M_{1}E}{\hbar ^{2}}. \label{parm}
\end{align}
We are not interested in situations with $M_{1}\leq 0$. As expected, the solution given by Eq. (\ref{solutionHC}) accommodates both symmetric and antissymmetric eigenstates.
In order to obtain the energies, we need to make use of the relations (\ref
{cndA}) and (\ref{cndB}). From condition (\ref{cndA}), we obtain
\begin{equation}
E=\hbar \Omega \left( 2n+\frac{1}{2}x+\frac{3}{2}\right) .
\label{Energy_condA}
\end{equation}
The second termination condition, Eq. (\ref{cndB}), for $n=0$, provides
\begin{equation}
\Omega =\frac{\hbar \omega ^{2}}{M_{1}}\left( \frac{x}{2}-m^{2}+\frac{2M_{1}E
}{\hbar ^{2}\omega ^{2}}+\frac{M_{1}}{M_{2}}+\frac{1}{2}\right).
\label{Frequency}
\end{equation}
Substituting Eq. (\ref{Frequency}) in Eq. (\ref{Energy_condA}) and then solving for $E$, we find
\begin{equation}
E_{0m}=\frac{\hbar ^{2}\omega ^{2}}{4M_{1}}\frac{x+3}{x+2}\left( 2m^{2}-x-
\frac{2M_{1}}{M_{2}}-1\right).
\end{equation}
The ground state energy has its minimum shifted as $M_{1}$ and $M_{2}$ are varied. The spacing between the levels increases as $m$ is increased. (Figure \ref{Energy_Ground_State}\,(a)).
When $M_{2}$ assumes ever lower positive values and $M_{1}$ is kept fixed (column (b) in the Table \ref{Table1}), the minimum energy state is not shifted. (Fig. \ref{Energy_Ground_State}\,(b)).
When $M_{1}$ is kept fixed and $M_{2}$ assumes negative values (column (c) in the Table \ref{Table1}), different from Fig. \ref{Energy_Ground_State}\,(a), there is no inversion between levels and, furthermore, the state with $m=4$ is not allowed (Fig. \ref{Energy_Ground_State}\,(c)). This characteristic is a consequence of the anisotropic mass. On the other hand, when $M_{1}=0.1$ is fixed and $M_{2}$ assumes negative values (column (d) in the Table \ref{Table1}), the inversion between levels is absent. However, only the energy with $m=1$ is allowed (Fig. \ref{Energy_Ground_State}\,(d)).

For the energy with $n=1$, the condition (\ref{cndB}) requires that
\begin{equation}
\det \left\vert
\begin{array}{cc}
\mu -q_{1} & 1+\beta  \\
\alpha  & \mu -q_{2}+\alpha
\end{array}%
\right\vert =0, \label{detn2}
\end{equation}
with
\begin{align}
q_{1} &=0, \\
q_{2} &=2+\beta +\gamma.
\end{align}
The solution of (\ref{detn2}) by using Eqs. (\ref{mu}), (\ref{nu}) and (\ref{parm}) provides two values for $\Omega$:
\begin{equation}
\Omega _{1}=\frac{1}{{M_{{1}}}\hbar }\left[  \hbar ^{2}{\omega
}^{2}\,X+\frac{6}{5}M_{{1}}E+\frac{2}{5}\sqrt{Y}\right] ,  \label{Omega1}
\end{equation}%
\begin{equation}
\Omega _{2}=\frac{1}{{M_{{1}}}\hbar }\left[ \hbar ^{2}{\omega }%
^{2}\,X+\frac{6}{5}\,M_{{1}}E-\frac{2}{5}\,\sqrt{Y}\right],  \label{Omega2}
\end{equation}%
with
\begin{equation}
X=\frac{1}{2}\,x-\frac{3}{5}%
\,{m}^{2}+\,\frac{3M_{1}}{5M_{2}}+{\frac{17}{10}},
\end{equation}
and
\begin{align}
Y &={\hbar }^{4}{m}^{4}{\omega }^{4}-2\,\frac{M_{1}}{M_{2}}{\hbar }^{4}{m}%
^{2}{\omega }^{4}+\left( \frac{M_{1}}{M_{2}}\right) ^{2}{\hbar }^{4}{\omega }%
^{4}-4\,{\hbar }^{4}{m}^{2}{\omega }^{4}\notag \\&+4\frac{M_{1}}{M_{2}}{\hbar }^{4}{%
\omega }^{4}+5\,{\hbar }^{4}{\omega }^{4}x  -4\,E{\hbar }^{2}{m}^{2}{\omega }^{2}M_{{1}}+14\,{\hbar }^{4}{\omega }%
^{4}\notag \\&+4\,E\frac{M_{1}^{2}}{M_{2}}{\hbar }^{2}{\omega }^{2}+8\,E{\hbar }^{2}{%
\omega }^{2}M_{{1}}+4\,{E}^{2}{M_{{1}}}^{2}.
\end{align}
However, after we replace (\ref{Omega1}) and (\ref{Omega2}) in (\ref{Energy_condA}) and solve the resulting equation for $E$, we find equal spectra. Thus, we write
\begin{equation}
E_{1}=\frac{{\hbar }^{2}{\omega }^{2}\left( 3+x\right) }{4M_{{1}}\left( 4-{x}%
^{2}\right) }\,\left(Q-2\,\sqrt{%
W}\right),  \label{Om1E1}
\end{equation}%
\begin{equation}
E_{2}=\frac{{\hbar}^{2}{\omega }^{2}\left( 3+x\right)}{4M_{{1}}\left( 4-{x}^{2}\right) }\,\left(Q+2\,\sqrt{W}\right), \label{Om1E2}
\end{equation}%
with
\begin{align}
Q &= 3\,{x}^{2}-2\,{m}^{2}x+2\,x\frac{M_{1}}{M_{2}}+11\,x+4, \\
W &=4\,{m}^{4}-4\,{m}^{2}{x}^{2}+{x}^{4}-8\,\frac{M_{1}}{M_{2}}{m}^{2}+4%
\frac{M_{1}}{M_{2}}{x}^{2}\notag \\
&+12\,{x}^{3}-16\,{m}^{2}x+4\,\left( \frac{M_{1}}{M_{2}}\right) ^{2}+16\,\frac{M_{1}}{M_{2}}x\notag \\&+38\,{x}^{2}-28\,{m}
^{2}+28\,\frac{M_{1}}{M_{2}}+40\,x+17.
\end{align}
The energy eigenvalues (\ref{Om1E1}) and (\ref{Om1E2}) present some exotic characteristics manifested by both isotropy and anisotropy in the masses.
In the isotropic case, all quantum numbers $m$ are allowed (Figure  \ref{Fig_2D_Energy_n1}\,(a)). The energies (\ref{Om1E1}) are responsible for the largest number of negative states while (\ref{Om1E2}) represents only positive states (Figures  \ref{Fig_2D_Energy_n1}\,(b)-(d)). The isotropic mass also reveals that energy states with $m=3$ are nearer when compared to others. In the isotropic case, the relation (\ref{Om1E1}) holds the largest number of states with negative energy while (\ref{Om1E2}) exhibits only positive energy values. This situation is just the opposite of what occurs in the isotropic case. Independent of the values of $M_{1}$ and $M_{2}$, the anisotropic mass affects the full energy states, so that some values of $m$ are not allowed (Figures  \ref{Fig_2D_Energy_n1}\,(b)-(d)). This characteristic is typical one of the so-called localized energy states. The appearing of these states is also a consequence of the limitation imposed by the relation $x=\sqrt{4M_{1}/M_{2}+1}$ (defined in Eq. (\ref{parm})), which imposes restrictions on the values of $M_{1}$ and $M_{2}$ to guarantee that $4M_{1}/M_{2}+1 \geq 0$.
Besides, we can note that by making some particular choices for the effective masses, we can change how the energy levels are filled. It means, in principle, that these choices could be used to obtain a specific energy profile, filtering the states by their angular momentum $m$. Thus, the geometry and the anisotropy can produce a combined effect in changing the form of the occupation levels.

\section{Conclusions}
\label{sec:conclusions}

In the present manuscript, we have addressed the motion of a quantum particle on a helicoidal geometry. We have considered the well-established formalism of the geometric potential in the context of anisotropic masses. From the equation for a quantum particle on a helicoid, we have inserted an harmonic oscillator potential into the effective potential through the vector coupling. Then, we calculated the eigenfunctions and energy eigenvalues. These energy eigenvalues are obtained from non-trivial relations since the wavefunction depends on the Confluent Heun Function.
In our case, the form of the solution does not allow a negative effective mass $M_1$, while $M_2$ can be either positive or negative. Despite that, it was possible to consider different configurations with respect to the values of the effective masses. In the case in which both masses are positive, configurations with arbitrary values for $M_1$ and $M_2$ are possible. However, in the case with $M_2 < 0$, it is necessary to be careful: the solution requires $x \geq 0$. Thus, in the case of an electronic analog of a hyperbolic material in the presence of an harmonic oscillator potential, the anisotropy tends to be larger than in the case of a positive mass $M_2$.
The system can exhibit several different behaviors, depending on the adjustment between the values of the masses. For instance, we have noted that the effective potential can be totally modified by making particular choices for effective masses. In addition, states with different values of angular momentum $m$ can be affected differently when subjected to the same type of effective potential. The characteristics observed in the sketch of the effective potential as well as in the energy spectrum allow us to apply this model to others systems in the domain of nanoscale physics, as for example, rings and quantum dots.

\section*{Acknowledgments}

This work was partially supported by the Brazilian agencies CAPES, CNPq and
FAPEMA. EOS acknowledges CNPq Grants 427214/2016-5 and 303774/2016-9, and
FAPEMA Grants 01852/14 and 01202/16. MMC acknowledges Coordena\c{c}\~{a}o de
Aperfei\c{c}oamento de Pessoal de N\'{\i}vel Superior - (CAPES) - Brasil - Grant
88887.358036/2019-00.

\bibliographystyle{apsrev4-2}

\end{document}